\documentclass[12pt]{article}
\usepackage{epsfig}
\usepackage{a4}
\usepackage{axodraw}
\usepackage{pslatex}

\begin{document}
\begin{center}
{\huge \bf The reaction $e^-e^+\rightarrow h h $ recomputed.
} \\[8mm]
{ \Large
 J.J. Lopez-Villarejo$^a$, J.A.M. Vermaseren$^b$ \\ [3mm]
}
\begin{minipage}{8cm}
\begin{itemize}
\item[$^a$]
Departamento de F\'\i sica Te\'orica, C-XI, \hfill \\
Universidad Aut\'onoma de Madrid, \hfill \\
Canto Blanco, \hfill \\
E-28034 Madrid, SPAIN.
\item[$^b$]
 NIKHEF, P.O. Box 41882, \hfill \\
 1009 DB, Amsterdam, The Netherlands
\end{itemize}
\end{minipage} \vspace{5mm}
\end{center}

\begin{abstract}
We notice that the existing literature about the reaction 
$e^-e^+\rightarrow h h$ suffers from a mistake in the relative sign between 
the t-channel and u-channel graphs. Correcting this lowers the crosssections 
by about an order of magnitude.
\end{abstract}

\newpage

\section{Introduction}

The reaction $e^-e^+\rightarrow h h$ is an interesting one from the 
theoretical viewpoint. Because the coupling of the higgs to electrons is 
proportional to the electron mass, the tree graphs are so small that the 
one loop contribution gives the dominant part of the crosssection. This has 
been recognized in the past~\cite{Hoogeveen,Djouadi} and it has been found 
that the crosssections are rather small. Only for special higgs particles 
in for instance SUSY models can the crosssection be larger, mainly due to a 
bigger tree graph contribution. In principle this would finish the subject 
if it were not for the fact that during a project in which we were testing 
methods we recalculated this reaction and found both papers to have a 
number of shortcomings. Correction of these altered the final numbers 
significantly. 

Of course this doesn't mean that now this reaction will become important at 
a future ILC, but at least reports about the variety of reactions that may 
or may not be of interest can quote the correct numbers. And if there exist 
extra higgs particles the errors of the past can be avoided in the future.

The major problem with one loop calculations is that they involve much 
work. Hence we like to use symmetry arguments as much as possible to 
minimize the number of diagrams to compute. A traditionally safe symmetry 
to use is either Bose or Fermi statistics when there are identical 
particles involved. Because there are two higgs particles involved, both 
previous papers have used this symmetry, but both got the relative sign of 
the diagrams wrong. This is because they used an indirect argument that for 
identical particles the amplitude should be maximal at the region of 
overlap and the scattering at 90 degrees was seen as the region of overlap. 
They adjusted the relative sign for this. We will show in the next section 
that this is not the case and that actually at 90 degrees the amplitude is 
zero. This diminishes the crosssection considerably. We have verified this 
result with the use of the FeynArts~\cite{FeynArts} system and it agrees 
with our findings.

This letter is arranged as follows. We start with explaining the relative 
sign of the graphs in the formulas, using the tree graphs as an example. 
Then we show the diagrams to compute and how the sign of the tree graphs 
transfers to the loop graphs. Because the calculation itself can be done in 
many ways and can even be done automatically with systems like FeynArts, we 
don't go into unnecessary details. Finally we present the new numbers and 
show a few distributions.

\section{The calculation}

The lowest order of the reaction $e^-e^+\rightarrow h h$ has in principle 
three diagrams, which we will label the t-channel, the u-channel and the 
s-channel graphs.
%
%
\begin{center}
\begin{picture}(310,60)(0,0)
%
%
  \ArrowLine(20,50)(50,50)
  \ArrowLine(50,50)(50,10)
  \ArrowLine(50,10)(20,10)
  \DashLine(50,50)(80,50){3}
  \DashLine(50,10)(80,10){3}
  \Vertex(50,50){1.5}
  \Vertex(50,10){1.5}
  \Text(13,50)[]{$p_a$}
  \Text(13,10)[]{$p_b$}
  \Text(89,50)[]{$p_1$}
  \Text(89,10)[]{$p_2$}
  \Text(45,30)[r]{$t$}
\Text(100,30)[]{$+$}
%
%
  \ArrowLine(120,50)(150,50)
  \ArrowLine(150,50)(150,10)
  \ArrowLine(150,10)(120,10)
  \DashLine(150,50)(180,10){3}
  \DashLine(150,10)(180,50){3}
  \Vertex(150,50){1.5}
  \Vertex(150,10){1.5}
  \Text(113,50)[]{$p_a$}
  \Text(113,10)[]{$p_b$}
  \Text(189,50)[]{$p_1$}
  \Text(189,10)[]{$p_2$}
  \Text(145,30)[r]{$u$}
\Text(200,30)[]{$+$}
%
%
  \ArrowLine(220,50)(240,30)
  \ArrowLine(240,30)(220,10)
  \DashLine(240,30)(270,30){3}
  \DashLine(270,30)(290,10){3}
  \DashLine(270,30)(290,50){3}
  \Vertex(240,30){1.5}
  \Vertex(270,30){1.5}
  \Text(213,50)[]{$p_a$}
  \Text(213,10)[]{$p_b$}
  \Text(299,50)[]{$p_1$}
  \Text(299,10)[]{$p_2$}
  \Text(255,25)[t]{$s$}
\end{picture}
\\
Fig. 1: The three tree graphs for the reaction $e^-e^+\rightarrow hh$
\end{center}

Because of the very small coupling of the higgs to the electron these 
diagrams are suppressed by powers of $(m_e/M_W)^2$ which is a number that 
is less than $10^{-10}$. The t-channel and u-channel graphs have two of 
such powers and the s-channel graph has only one. Let us however, for study 
purposes, look at the t-channel and u-channel graphs assuming that the 
coupling is a fixed constant c that doesn't depend on the electron mass. In 
that case the amplitude for the t-channel graph becomes:
\begin{eqnarray}
	A_t & = & c^2\ \overline{v}(p_b) \frac{\gamma_\mu p_a^\mu
		-\gamma_\mu p_1^\mu+m_e}{t-m_e^2} u(p_a)
\end{eqnarray}
Using the Dirac equation in the limit that the electron mass goes to zero 
this gives
\begin{eqnarray}
	A_t & = & -c^2\ \frac{1}{t}\ \overline{v}(p_b)\ \gamma_\mu p_1^\mu\ u(p_a)
\end{eqnarray}
For the u-channel diagram we have to exchange $p_1$ and $p_2$ and hence 
also $t$ and $u$ which gives
\begin{eqnarray}
	A_u & = & -c^2\ \frac{1}{u}\ \overline{v}(p_b)
		\ \gamma_\mu p_2^\mu\ u(p_a) \nonumber \\
	    & = & +c^2\ \frac{1}{u}\ \overline{v}(p_b)
		\ \gamma_\mu p_1^\mu\ u(p_a)
\end{eqnarray}
in which we replaced $p_2$ by $p_a+p_b-p_1$ and the $p_a$ and $p_b$ can be 
removed with the Dirac equation. It is this change in sign, which at first 
looks counter intuitive, that caused all the problems because this way the 
sum of the t-channel and u-channel graphs becomes proportional to $t-u$:
\begin{eqnarray}
	A_t+A_u & = & c^2\ \frac{t-u}{tu}
	\ \overline{v}(p_b)\ \gamma_\mu p_1^\mu\ u(p_a)
\end{eqnarray}
and the result is that at 90 degrees this part of the amplitude vanishes. 
The complete matrix element is still invariant under the exchange of $t$ 
and $u$ as $(t-u)^2$ and $(tu)$ are invariant under this exchange and all 
$t$ and $u$ dependence can be written in terms of these two variables. 
Their contribution to the full matrix element becomes:
\begin{eqnarray}
	|A_t+A_u|^2 & = & 2\ c^4\frac{(t-u)^2}{tu}(1-\frac{m_h^2}{tu})
\end{eqnarray}
We leave out the s-channel graph because the equivalent one loop graphs 
are not important: they always vanish in the limit that the electron mass 
is put to zero.

Let us see how this works out at the one loop level.
%
%
\begin{center}
\begin{picture}(280,60)(0,0)
%
%
  \ArrowLine(20,50)(50,50)
  \ArrowLine(50,50)(50,10)
  \ArrowLine(50,10)(20,10)
  \Photon(50,50)(90,50){3}{4.5}
  \Photon(90,50)(90,10){3}{4.5}
  \Photon(90,10)(50,10){3}{4.5}
  \DashLine(90,50)(120,50){3}
  \DashLine(90,10)(120,10){3}
  \Vertex(50,50){1.5}
  \Vertex(50,10){1.5}
  \Vertex(90,50){1.5}
  \Vertex(90,10){1.5}
  \Text(13,50)[]{$p_a$}
  \Text(13,10)[]{$p_b$}
  \Text(129,50)[]{$p_1$}
  \Text(129,10)[]{$p_2$}
  \Text(70,45)[t]{$W$}
  \Text(85,30)[r]{$W$}
  \Text(70,15)[b]{$W$}
%
%
  \ArrowLine(160,50)(190,50)
  \ArrowLine(190,50)(190,10)
  \ArrowLine(190,10)(160,10)
  \Photon(190,50)(230,50){3}{4.5}
  \DashLine(230,50)(230,10){3}
  \Photon(230,10)(190,10){3}{4.5}
  \DashLine(230,50)(260,50){3}
  \DashLine(230,10)(260,10){3}
  \Vertex(190,50){1.5}
  \Vertex(190,10){1.5}
  \Vertex(230,50){1.5}
  \Vertex(230,10){1.5}
  \Text(153,50)[]{$p_a$}
  \Text(153,10)[]{$p_b$}
  \Text(269,50)[]{$p_1$}
  \Text(269,10)[]{$p_2$}
  \Text(210,45)[t]{$W$}
  \Text(227,30)[r]{$X$}
  \Text(210,15)[b]{$W$}
\end{picture}
\\
Fig. 2: Some one loop graphs in the reaction $e^-e^+\rightarrow hh$
\end{center}
There are two types of diagrams: the ones with the vector bosons and the 
ones with the ghosts. As it turns out, the ones with the vector bosons are 
the easier ones as they involve only one power of the loop momentum in the 
numerator. Due to the Dirac equation and the fact that the loop momentum 
gives us, after integration, only Lorenz tensors that involve the external 
momenta only the box diagrams are nonzero in the limit that the electron 
mass becomes zero. If the terms in a diagram have only an odd number of 
gamma matrices ($\gamma_5$ counts as an even number for these purposes), 
eventually the diagram becomes of the type
\begin{eqnarray}
	A & = & \overline{v}(p_b)\ \gamma_\mu p_1^\mu\ (F_1(s,t,u)
	 + F_2(s,t,u)\gamma_5)\ u(p_a) 
\end{eqnarray}
and when exchanging $(p_1,t,u)$ and $(p_2,u,t)$ we get the same minus sign 
as before. This is different for an even number of gamma matrices where we 
have a relative plus sign. In the case of the standard model however we 
have only diagrams with an odd number of gamma matrices when we put the 
electron mass equal to zero.

While doing the calculation we ran into another problem with 
reference~\cite{Hoogeveen}. Their figures show a rather strange and 
unphysical behaviour when the higgs mass is varied. We could track this 
down to the exchange of two indices in a tensor in their equation 8. Once 
this is corrected we get a much more physical behaviour although we still 
don't get exactly the same result. There are however not enough details to 
investigate this further. We don't understand the statement in 
reference~\cite{Djouadi} that they agree with the previous paper as they 
seem to have only a problem with the relative sign of the diagrams.

For the rest the calculation was rather standard. We evaluated the standard 
model higgs reaction first using FORM~\cite{FORM} to reduce to scalar loop 
integrals according to the methods in ref~\cite{Oldenborgh}. Then we did 
the scalar loop integrals by standard techniques. Eventually we also used 
FeynArts as an extra check as one should take disagreement with two papers 
not lightly. We would also have used the GRACE~\cite{GRACE} system as a 
check, but unfortunately the version we had wasn't ready for it yet and for 
the version that might have handled it we didn't have the proper supporting 
programs.
%
%

\section{Results}

Because in the old calculations the amplitude reached a maximum at 90 
degrees and the new calculation gives zero there, it should come as no 
surprise that the total crosssection is now significantly smaller. We give 
some numbers:
\begin{center}
\begin{tabular}{|c|c|c|}
	\hline
   CM energy(GeV) & Higgs Mass(GeV) & Crosssection(fb) \\ \hline
         500      &        120      &        0.0157        \\
         500      &        150      &        0.0116        \\
        1000      &        120      &       0.00964        \\
        1000      &        150      &       0.00924        \\
        1000      &        200      &       0.00791        \\
        1000      &        300      &       0.00439        \\ \hline
\end{tabular}
\vspace{2mm} \\
Corrected crosssections for the reaction $e^-e^+\rightarrow hh$ in the 
standard model.
\end{center}

In the table we show a number of crosssections for different energies and 
masses. This can be used as a reference to test programs. For the coupling 
constants we used the running values. This is not automatic in the FeynArts 
system.

\begin{center}
\epsfig{file=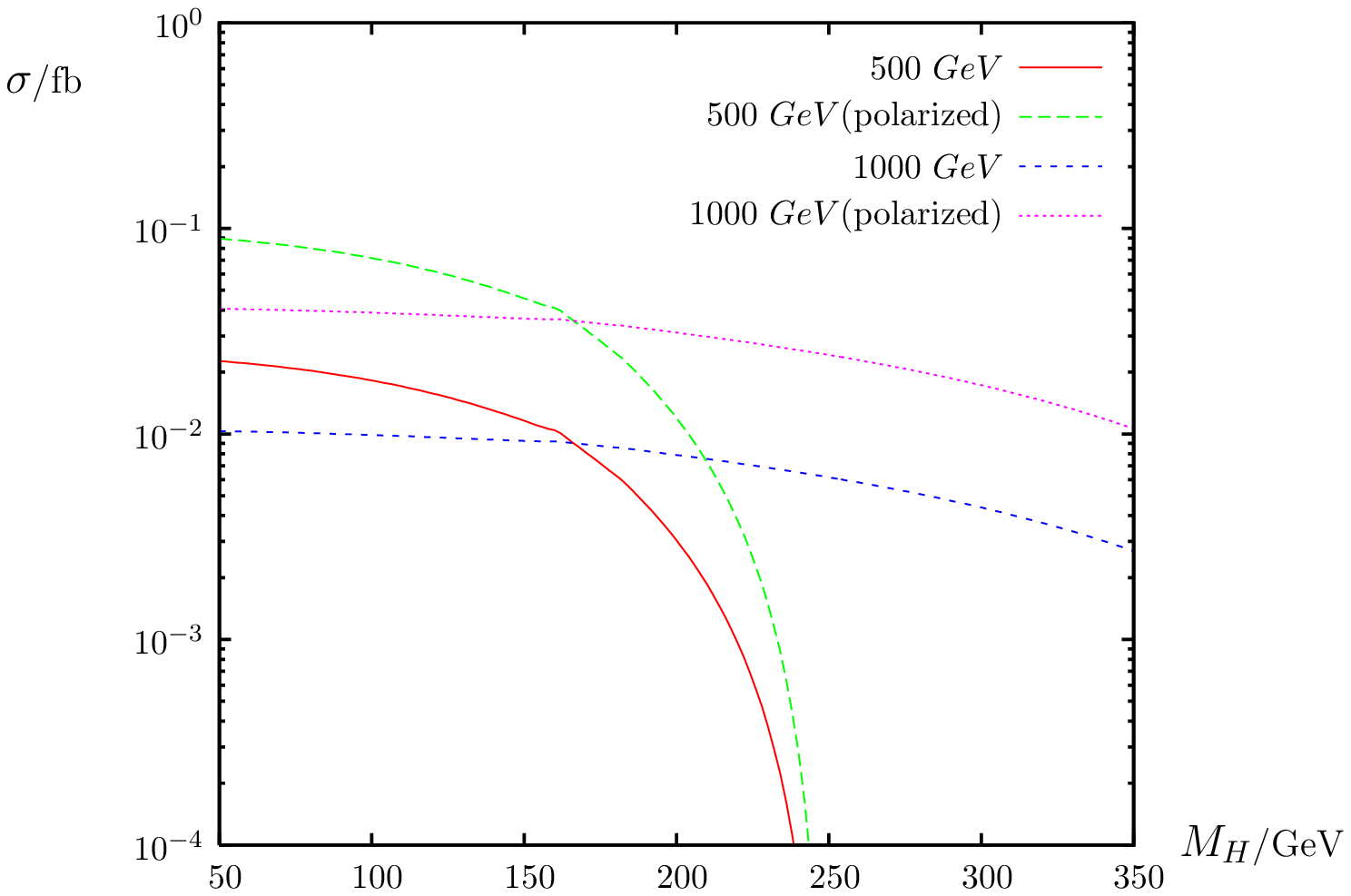,width=12cm,angle=0} \\
Fig. 3: The crosssection as a function of the mass of the higgs.
\end{center}

In figure 3 we present the crosssections as a function of the mass of the 
higgs. We notice that for purely polarized beams we can get a four times 
larger crosssection.

\begin{center}
\epsfig{file=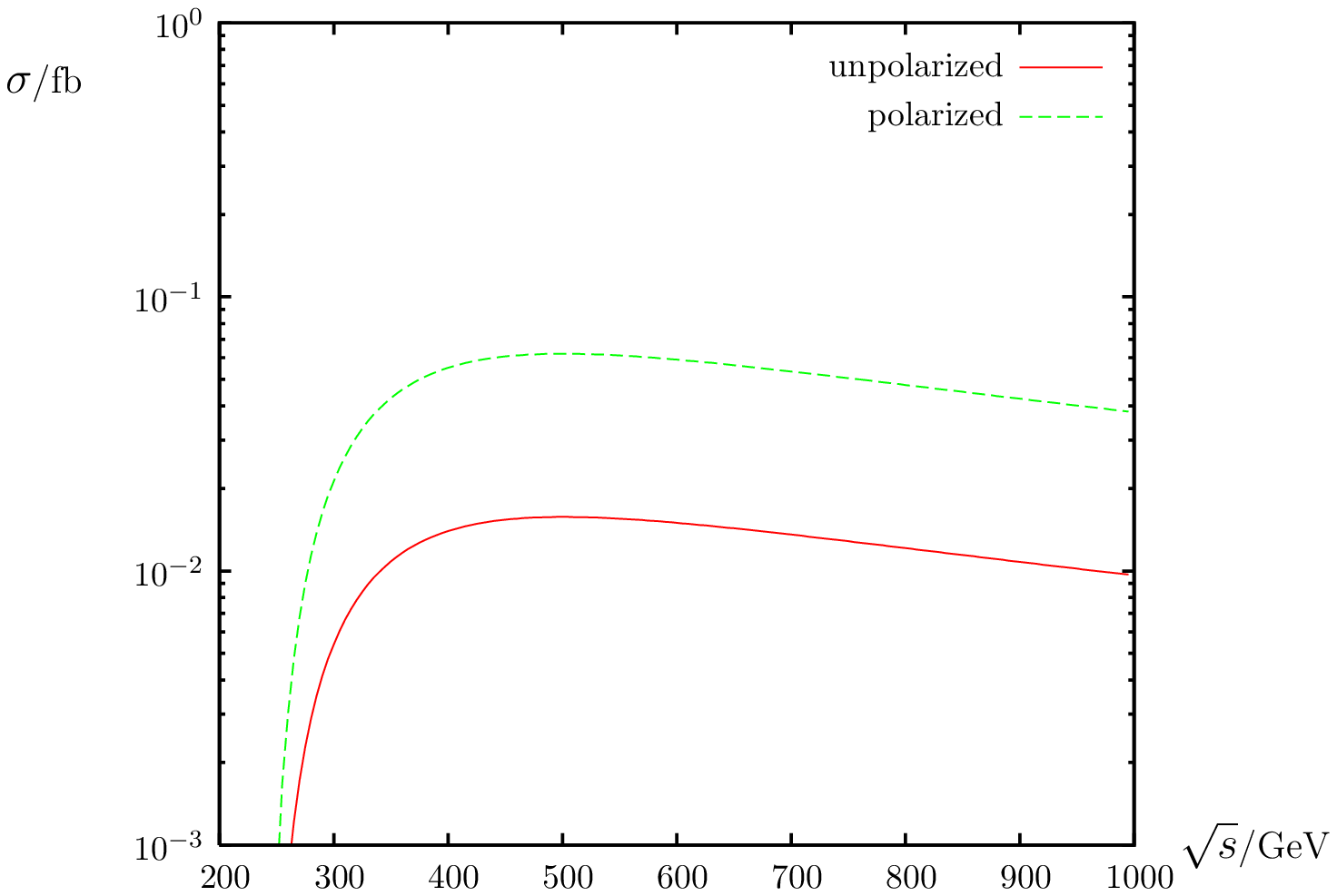,width=12cm,angle=0} \\
Fig. 4: The polarized and unpolarized crosssection as a function of the CM energy. 
The mass of the higgs is 120 GeV.
\end{center}

In figure 4 we show the crosssections as a function of the CM energy for a 
120 GeV higgs mass. We see a slowly declining crosssection when we are away 
from the threshold. This must clearly be a loop effect, as the tree graphs 
would show a slow (logarithmic) rise.

\begin{center}
\epsfig{file=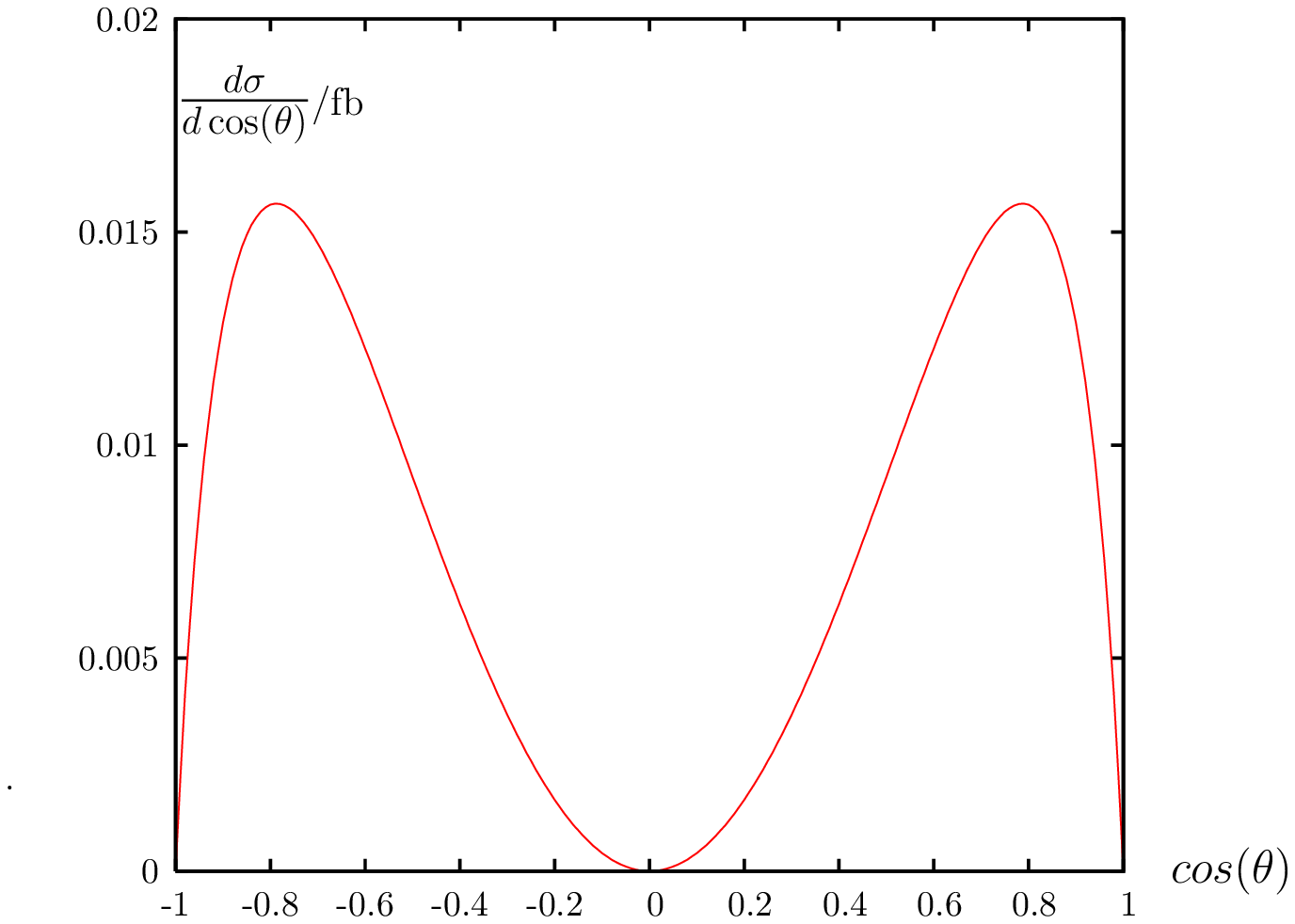,width=12cm,angle=0} \\
Fig. 5: The differential crosssection w.r.t. theta in the CM frame. 
The mass of the higgs is 120 GeV and the total CM energy is 500 GeV.
\end{center}

In figure 5 we see the differential distribution in terms of the cos of the CM 
azimuthal angle. As mentioned before, this distribution should be zero at 
90 degrees when $\cos\theta=0$. The immediate consequence of this is that 
if the energy is much above threshold, one should look for these events in 
a different region of phase space than the one that the previous papers 
would indicate.

We have also redone some of the MSSM calculations of ref~\cite{Djouadi}. 
For this we used solely the FeynArts system~\cite{FeynArts,FeynArtsMSSM}. 
In this case the situation is potentially more complicated, but the 
diagrams with even numbers of gamma matrices between the spinors are much 
smaller again and hence the same sign problem makes that the crosssections 
becomes correspondingly smaller than in the original calculation. This 
means that the ratio between the signals from the standard model and the 
MSSM model doesn't change significantly.
 
The first author acknowledges financial support from the Comunidad de Madrid 
through a FPI contract and the helpful advice of T. Hahn regarding the FeynArts
system. The work of second author was supported by FOM and the Humboldt foundation.
We would like to thank DESY Zeuthen for its hospitality during part of this work.

\end{document}